\RequirePackage{fix-cm}
\documentclass[]{relaxed_system_lab}

\ifPDFTeX
\usepackage[utf8]{inputenc}
\fi
\usepackage[T1]{fontenc}
\usepackage{charter}
\usepackage{xspace,flafter,placeins}
\usepackage{needspace}
\usepackage{amsmath,amssymb}
\usepackage{graphicx,booktabs,tabularx,array}
\usepackage{wrapfig} 
\usepackage{xcolor,microtype,xurl}
\usepackage{enumitem}
\usepackage{algorithm,algorithmic}

\newcommand{\sys}{$e$\textsc{FAIR}\xspace}

\newif\ifdraftnotes
\draftnotestrue
\newif\ifappendixdetails
\appendixdetailsfalse

\newcolumntype{Y}{>{\raggedright\arraybackslash}X}

\newcommand{\proc}[1]{\textsc{#1}}

\appendixdetailstrue 

\hypersetup{
  pdftitle={eFAIR: Fair LLM Serving under Hourly Electricity Commitments},
  pdfauthor={You Peng, Youhe Jiang, Chen Wang, Binhang Yuan}
}
\title{\sys: Fair LLM Serving under Hourly Electricity Commitments}
\author{You Peng$^1$$^*$, Youhe Jiang$^1$$^*$, Chen Wang$^1$, Binhang Yuan$^1$}
\affiliation{$^1$HKUST}
\contribution{$^*$Equal contribution}

\abstract{Datacenters can procure electricity in day-ahead markets by committing to energy purchases for individual hours.
For LLM services operating within these hourly budgets, maximizing aggregate throughput alone may still leave individual requests waiting or receiving much slower service than desired.
Fair serving therefore requires improving service for poorly served requests while reserving electricity for later arrivals whose timing and demand are unknown.
To compare service across requests, we define normalized throughput as each request's actual output rate divided by the product of its standalone throughput and its service entitlement, a positive weight representing its desired service level.
Using this measure, we formulate fair serving as leximin optimization under an hourly electricity constraint, prioritizing requests with the lowest normalized throughput.
To address this problem, we present \sys, an online scheduler that coordinates request dispatching, electricity budget allocation, and GPU frequency selection through a common normalized-throughput target.
Specifically, request dispatching prioritizes requests by desired service level and progress; electricity budget allocation reserves electricity for current and expected future work; and frequency selection uses electricity savings to raise GPU frequencies for requests with higher desired service levels.
Evaluations on H20 and H200 configurations, each with $32$ replicas, show that \sys improves the tenth percentile of normalized throughput by $28.0\%$--$31.0\%$ over the best budget-feasible baseline, while keeping absolute deviations from the hourly electricity commitment below $0.26\%$.
\sys also improves throughput balance among concurrent requests and aligns electricity allocation more closely with service demand over time.
Together, these results demonstrate \mbox{how \sys improves request-level fairness within an hourly commitment.}}

\begin{document}
\maketitle

\section{Introduction}
\label{sec:introduction}
Large language model (LLM) serving must distribute computing capacity among competing requests while controlling electricity consumption~\cite{sheng2024vtc,stojkovic2025dynamollm}.
In this paper, we ask: \textit{how can online LLM serving provide fair request throughput under hourly electricity commitments?}
This requires deciding both which requests to serve and when to spend the available electricity.
This question involves three key aspects.
(\underline{i}) \textbf{Online decisions:} The scheduler learns about individual requests only when they arrive and must make service decisions without knowing future arrivals.
(\underline{ii}) \textbf{Hourly electricity constraint:} An \textit{hourly electricity commitment} is an amount of electrical energy fixed before the hour; we treat this amount as an upper bound on the serving pool's total consumption during that hour.
Hourly procurement in day-ahead markets motivates this accounting period~\cite{ferc2024energyprimer,pjmEnergyMarkets,isoneDayAhead}.
(\underline{iii}) \textbf{Fair request throughput:} Fairness must account for differences in processing cost and desired service level.
We measure a request's throughput as the number of output tokens generated divided by the elapsed time from arrival, including waiting and execution.
Its \textit{standalone throughput}, i.e., the average output rate when served alone under a fixed reference configuration, provides a baseline for comparison.
In this paper, we define \textit{service entitlement} as a positive weight that scales the desired throughput relative to this baseline.
For requests with the same standalone throughput, an entitlement of 1.5 specifies a desired rate 1.5 times that specified by an entitlement of 1; it does not guarantee that rate.
We divide actual throughput by the product of standalone throughput and entitlement to obtain a \textit{normalized throughput} score, making service comparable across requests with different baselines and weights.
Subject to the hourly energy budget, our objective follows \textit{leximin fairness}~\cite{kurokawa2018leximin}: first maximize the lowest score, then the next lowest without reducing any previously maximized \mbox{score, and continue in ascending order.}

Prioritizing the lowest normalized-throughput scores under an hourly electricity budget is essential because a system can satisfy its electricity constraint and achieve high aggregate throughput while still providing poor service to individual requests.
A scheduler may process many tokens while leaving some requests waiting or serving them at rates far below those implied by their entitlements.
With a fixed hourly budget, energy spent on requests already receiving ample service reduces the amount available to improve those with lower scores.
Growing electricity demand from AI services strengthens the need to use available energy effectively~\cite{iea2025energyai,shehabi2024datacenter}.
A request-level fairness objective makes the distribution of that service explicit, \mbox{alongside total throughput and consumption.}

Two coupled difficulties prevent a straightforward allocation in our scenario.
First, requests have different processing costs and entitlements, so equal compute or energy shares need not produce equal normalized throughput.
Worker assignment, batch composition, and GPU frequency jointly determine both request progress and energy cost~\cite{patel2024power,hu2026ecoinfer}.
Prioritizing a request therefore requires execution choices that can translate priority into a higher service rate.
Second, the scheduler must allocate energy across arrival times without knowing when future requests will arrive.
Spending uniformly can restrict service during bursts, whereas spending only on visible work can exhaust the budget before later arrivals.
The scheduler must reserve energy for running iterations and expected future work, then release savings while they \mbox{can still improve service.}

Existing serving systems are insufficient to address the coupled difficulties of fair service allocation among concurrent requests and electricity allocation under uncertain future arrivals.
For example, vLLM improves serving throughput through efficient memory management and continuous batching~\cite{kwon2023pagedattention}.
Virtual Token Counter (VTC) balances cumulative client service using a configurable cost for processed tokens~\cite{sheng2024vtc}.
EcoInfer selects GPU frequencies to reduce energy consumption subject to latency objectives~\cite{hu2026ecoinfer}.
These mechanisms do not coordinate execution decisions based on individual requests' normalized throughput with electricity reservations for later arrivals, leaving fair allocation \mbox{across requests and time unresolved.}

We present \sys, an online scheduling heuristic that connects request priorities and electricity reservations through a common normalized-throughput target.
For concurrent requests, \sys coordinates worker assignment, batching, and GPU frequency according to entitlement and service progress, translating service priorities into shorter waiting times and faster execution.
Across arrival times, \sys adapts electricity allocation to observed demand and consumption, using electricity savings to improve current service while reserving electricity for expected later work.
Our contributions are summarized as follows:

\noindent\underline{\textbf{Contribution 1:}} We formulate fair LLM serving as leximin optimization of normalized throughput under an hourly electricity constraint (\S~\ref{sec:formulation}).
We identify three properties of fair serving: effective electricity use, balanced throughput among concurrent requests, and demand-aware electricity allocation.
Together, the objective and properties connect request fairness to the use of committed electricity.\looseness=-1

\noindent\underline{\textbf{Contribution 2:}} We develop \sys to coordinate service and electricity allocation (\S~\ref{sec:design}).
\proc{Request Dispatcher} prioritizes requests by entitlement and progress (\S~\ref{sec:service-dispatch}); \proc{Electricity Budget Allocator} maintains shared reservations (\S~\ref{sec:electricity-spending}); \proc{Frequency Selector} uses projected throughput to select execution configurations (\S~\ref{sec:frequency-choice}).
Together, these components direct the budget toward requests with weaker normalized service.\looseness=-1

\noindent\underline{\textbf{Contribution 3:}} We evaluate constrained fairness and all three properties on H20 and H200 configurations with 32 replicas each (\S~\ref{sec:evaluation}).
\sys improves the tenth percentile of normalized throughput by 28.0\%--31.0\% over the best budget-feasible baseline, with absolute consumption deviations below 0.26\%.
\sys also improves concurrent throughput balance and reduces the mismatch between demand and electricity consumption.\looseness=-1

\section{Background and Related Work}
\label{sec:preliminaries}
\label{sec:related}

We review three areas, starting with LLM systems that schedule requests and control inference costs (\S~\ref{sec:background-serving}).
We then discuss fairness based on resource shares and application performance in networks and training clusters (\S~\ref{sec:background-fairness}).
Finally, we review electricity procurement and workload scheduling over time (\S~\ref{sec:background-electricity}).
This progression connects request processing, capacity \mbox{sharing, and electricity consumption.}

\subsection{LLM Serving Systems}
\label{sec:background-serving}

\noindent\textbf{Continuous batching and KV-cache management.} LLM serving systems improve GPU utilization by processing multiple requests together.
Prefill processes the prompt, while decoding generates output tokens using the key and value (KV) cache.
Orca schedules requests at iteration boundaries, allowing completed requests to leave and new requests to enter a running batch~\cite{yu2022orca}.
vLLM reduces KV-cache fragmentation through PagedAttention, and Sarathi-Serve uses chunked prefill to limit interference with decoding~\cite{kwon2023pagedattention,agrawal2024sarathi}.
These mechanisms improve serving \mbox{capacity while controlling batching latency.}

\noindent\textbf{Request placement and preemption.} Request placement and preemption determine how that capacity translates into response times.
DistServe and Splitwise separate prefill and decoding across resources to accommodate the different requirements of the two phases~\cite{zhong2024distserve,patel2024splitwise}.
Llumnix migrates requests for load balancing and differentiated service, while FastServe uses iteration-level preemption to reduce request completion time~\cite{sun2024llumnix,wu2026fastserve}.
These systems manage where and when requests run to improve latency \mbox{under shared capacity}~\cite{jiang2023hexgen,jiang2025hexgen,jiang2025thunderserve,tong2025parallax,peng2026hexgen,peng2026hexagent}.

\noindent\textbf{Token-based client service accounting.} Client service counters guide admission when multiple clients compete for batch capacity.
Virtual Token Counter (VTC) measures cumulative client service using a configurable cost for processed input and output tokens~\cite{sheng2024vtc}.
VTC uses these counters to select requests for admission to continuous batches and bounds service differences between backlogged clients.
The weighted extension accounts for client entitlements by normalizing each client's \mbox{accumulated service by its assigned weight.}

\noindent\textbf{GPU provisioning and frequency scaling.} Inference systems adjust GPU capacity and operating frequency to satisfy power and performance requirements.
POLCA exploits measured inference power patterns to support power oversubscription~\cite{patel2024power}.
DynamoLLM adjusts instance count, model parallelism, and GPU frequency to reduce energy and cost subject to latency objectives~\cite{stojkovic2025dynamollm}.
EcoInfer predicts iteration latency and selects energy-efficient GPU frequencies under latency constraints~\cite{hu2026ecoinfer}.
These approaches connect serving configurations to power capacity and energy consumption~\cite{peng2026epact}.
\nocite{jiang2025demystifying,jiang2026boute}

\subsection{Fair Resource Allocation}
\label{sec:background-fairness}

\noindent\textbf{Network bandwidth sharing.} Network bandwidth is divided among competing flows according to their service shares.
Classical fair queueing and Start-Time Fair Queueing regulate service to backlogged flows, with weights specifying relative shares~\cite{demers1989fairqueueing,goyal1997sfq}.
Max-min fairness increases the smallest allocation first, subject to capacity constraints~\cite{jaffe1981bottleneck}.
Under weighted allocation, service is compared after division by the assigned weight.
The choice of service quantity and weights therefore determines which recipients \mbox{are considered underserved.}

\noindent\textbf{Resource shares and job performance.} Cluster allocation extends weighted sharing to multiple resource types and application-dependent performance.
Dominant Resource Fairness compares users by their largest share of any resource type~\cite{ghodsi2011drf}.
For GPU training jobs, Themis introduces finish-time fairness, while Gavel expresses scheduling policies through effective throughput on heterogeneous accelerators~\cite{mahajan2020themis,narayanan2020gavel}.
These approaches distinguish resource shares from application performance, since equal resource shares can \mbox{produce different progress rates.}

\subsection{Datacenter Electricity Management}
\label{sec:background-electricity}

\noindent\textbf{Hourly electricity procurement.} Electricity markets organize purchases and scheduled consumption over specified periods.
PJM's day-ahead market uses hourly prices, while hourly clean electricity matching aligns procurement with consumption in corresponding periods~\cite{pjmEnergyMarkets,epa2025hourlymatching}.
Electrical energy, measured in joules or kilowatt-hours, is the integral of power over time; a power limit constrains instantaneous power draw in watts.
The distinction determines whether management must limit instantaneous demand or allocate consumption across a settlement period.

\noindent\textbf{Energy-proportional computing and workload shifting.} Datacenter electricity demand depends on hardware utilization and when flexible workloads execute.
Energy-proportional computing seeks to reduce power as utilization falls, addressing substantial consumption at low load~\cite{barroso2007energyproportional}.
Carbon-aware computing shifts flexible workloads using hourly capacity limits while preserving daily processing capacity~\cite{radovanovic2023carbonaware}.
Hardware efficiency determines consumption during and between tasks, while workload scheduling determines when these costs occur.
Both affect the \mbox{electricity required for a given workload.}

\section{Problem Formulation}
\label{sec:formulation}

We formulate fair LLM serving as leximin optimization of normalized throughput under a fixed hourly electricity commitment.
\S~\ref{sec:problem-overview} first states the constrained scheduling problem and the information available during online serving.
\S~\ref{sec:fair-serving} defines normalized throughput, its service target, and the measurement at the hour endpoint.
\S~\ref{sec:commitment} then uses that target to set the commitment and combines the electricity constraint with the complete fairness objective.
Finally, \S~\ref{sec:properties} proposes three properties of desirable solutions.\looseness=-1

\subsection{Problem Overview}
\label{sec:problem-overview}

\noindent\textbf{Problem inputs.}
Online scheduling allocates service under a predetermined hourly electricity commitment as requests arrive.
The problem inputs include the commitment $C_h>0$ for hour $h$, a normalized-throughput target, and the associated demand estimate; \S~\ref{sec:commitment} explains how the target and demand estimate determine the commitment.
Resource limits $\mathcal R$ specify the replica count, memory capacity, and batch limits, while $\mathcal F$ denotes the set of available GPU frequencies.
Each request reveals a prompt and entitlement at arrival, while output progress and completion are observed during service.
Scheduling decisions therefore use these inputs and the observed serving \mbox{history without knowing future arrivals.}

\noindent\textbf{Scheduling decisions and objective.}
The objective is to improve the weakest request service within the hourly electricity commitment.
A scheduling policy $\pi$ assigns arriving requests to workers and batches and selects a GPU frequency for each batch.
Let $\Pi$ denote the set of online policies that obey the serving pool's resource limits, and $E_h(\pi)$ the hourly electricity consumption under policy $\pi$.
We next define request fairness (\S~\ref{sec:fair-serving}), then specify the electricity budget and complete objective (\S~\ref{sec:commitment}).
At a high level, the commitment constrains feasible allocations, while fairness determines the preferred allocation:
\begin{equation}
\begin{aligned}
\underset{\pi\in\Pi}{\operatorname{maximize}}\quad
&\operatorname{Fairness}(\pi)\\
\text{subject to}\quad &E_h(\pi)\le C_h.
\end{aligned}
\label{eq:abstract-problem}
\end{equation}

\subsection{Fair LLM Serving}
\label{sec:fair-serving}

Fair LLM serving normalizes each request's throughput by the product of its standalone throughput and entitlement, then prioritizes requests with the lowest normalized throughput.
We first define this measure and its service target, then specify how to compare all requests at the hour endpoint.\looseness=-1

\noindent\textbf{Normalized throughput.}
Normalized throughput accounts for both the effect of contention and the request's service entitlement.
Let $B_i>0$ denote request $i$'s \emph{standalone throughput}, the average output rate when request $i$ is served alone under a fixed reference configuration.
Let $w_i>0$ be the request's entitlement, $N_i$ the output token count, and $a_i,c_i$ the arrival and completion times.
For completed requests, actual and normalized throughput are
\begin{equation}
T_i=\frac{N_i}{c_i-a_i},\qquad q_i=\frac{T_i}{w_iB_i}.
\label{eq:q}
\end{equation}

\Needspace{255pt}
\begin{wrapfigure}{r}{0.47\textwidth}
\vspace{-5pt}
\centering
\captionsetup{justification=raggedright,singlelinecheck=false}
\includegraphics[width=\linewidth,trim=0 5pt 0 4pt,clip]{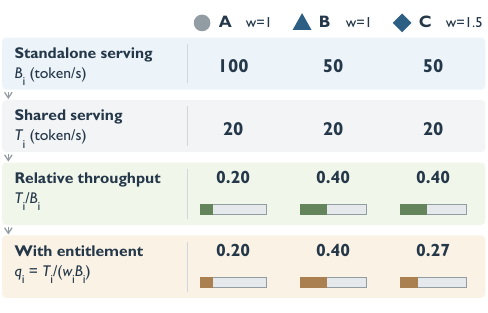}
\caption{\begingroup\parfillskip=0.4\linewidth minus 0.4\linewidth\relax Normalization for three requests.
Equal actual throughput of 20 tokens per second yields relative throughputs of 0.20, 0.40, and 0.40; dividing by entitlement gives normalized throughputs \mbox{of 0.20, 0.40, and 0.27.}\endgraf\endgroup}
\label{fig:normalization}
\vspace{-3pt}
\end{wrapfigure}
Queueing, prefill, and decoding all count toward duration, so comparable $q_i$ values represent comparable service relative to standalone throughput and entitlement.
Figure~\ref{fig:normalization} shows why equal actual throughput can yield different normalized throughput.
A common target $z>0$ specifies the desired normalized-throughput floor: $q_i\ge z$ means $T_i\ge zw_iB_i$, or at least $zw_i$ times the request's standalone throughput.
This interpretation connects protection of weak requests to the service demand used in commitment planning (\S~\ref{sec:commitment}) and guides online allocation.\looseness=-1

\noindent\textbf{Hour-end normalized throughput.}
We assess all requests arriving in the hour at the common endpoint $H$.
Let $\mathcal I_h$ be this request set, $n_i(H)$ the output produced by $H$, and $\tau_i^H$ the time from arrival to completion, or to $H$ for an unfinished request.
Observed normalized throughput is $q_i^H=\frac{n_i(H)}{w_iB_i\tau_i^H}$, with zero output giving $q_i^H=0$.
The standalone throughput $B_i$ remains fixed for each request across scheduling policies; the appendix specifies the endpoint rules.
This measurement includes waiting and unfinished requests when comparing service within the hour.
With request service defined, we next specify the electricity constraint and the resulting fairness objective.\looseness=-1

\subsection{Fair Serving under Electricity Commitments}
\label{sec:commitment}

The hourly commitment bounds the electricity available for fair serving.
We use the service target from \S~\ref{sec:fair-serving} to plan the budget, define consumption and commitment deviation, and combine these quantities with hour-end normalized throughput in the complete optimization problem.\looseness=-1

\noindent\textbf{Setting the hourly commitment.}
The operator sets $C_h$ from the expected workload, entitlement mix, and normalized-throughput target $z$ from \S~\ref{sec:fair-serving}.
Historical workloads provide an estimate of the requests to be served; the electricity required to support their target rates $zw_iB_i$ gives the aggregate service-electricity estimate $\widehat D_h$.
Let $\Delta_h$ be the hour duration and $P_{\mathrm{floor}}$ the declared fixed operating power.
The commitment is $C_h=P_{\mathrm{floor}}\Delta_h+\widehat D_h$.
The scheduler receives $\widehat D_h$ with $C_h$ to reserve electricity for later arrivals while serving current work; the appendix specifies the common demand-estimation rule.
The commitment fixes the hourly electricity budget, while scheduling determines when electricity is spent and which requests receive service.\looseness=-1

\noindent\textbf{Measuring commitment deviation.}
Commitment deviation compares total hourly consumption with the planned budget.
Consumption $E_h$ accounts for active execution, idle GPUs, and fixed operating costs.
The signed deviation is $\delta_h=\frac{E_h-C_h}{C_h}$ and the absolute deviation is $d_h=|\delta_h|$.
Positive $\delta_h$ denotes overuse and negative $\delta_h$ denotes underuse; a policy is feasible in Equation~\ref{eq:abstract-problem} when $E_h\le C_h$.
This accounting determines budget feasibility and quantifies how closely consumption matches the commitment.\looseness=-1

\noindent\textbf{Leximin objective.}
Within this budget, leximin fairness improves the weakest observed service first.
Given the hourly electricity commitment $C_h$, normalized-throughput target $z$, associated demand estimate $\widehat D_h$, resource limits $\mathcal R$, and frequency set $\mathcal F$, let $\mathbf q_h^{\uparrow}(\pi)$ contain the normalized throughputs $q_i^H(\pi)$ for requests in $\mathcal I_h$, sorted in nondecreasing order.
With $q_i^H$ defined in \S~\ref{sec:fair-serving} and $E_h$ \mbox{accounted for above, the complete problem is}
\begin{equation}
\begin{aligned}
\underset{\pi\in\Pi(\mathcal R,\mathcal F)}{\operatorname{lexmax}}\quad
&\mathbf q_h^{\uparrow}(\pi)\\
\text{subject to}\quad &E_h(\pi)\le C_h,\\
&q_i^H(\pi)=\frac{n_i(H;\pi)}{w_iB_i\tau_i^H(\pi)},
\quad i\in\mathcal I_h.
\end{aligned}
\label{eq:fair-serving-objective}
\end{equation}

Here $\Pi(\mathcal R,\mathcal F)$ is the online policy set from \S~\ref{sec:problem-overview}, restricted to the stated resource limits and frequencies.
The leximin objective first maximizes $Q_{\min}=\min_i q_i^H$, then the next lowest normalized throughput without reducing any previously maximized value, and continues through the sorted vector.
Policies with the same minimum can therefore differ in the service received by the remaining weak requests.
This objective defines the fairness goal; the following properties characterize desirable solutions under the commitment.

\subsection{Properties of Fair Serving}
\label{sec:properties}

We propose three properties for assessing solutions to Equation~\ref{eq:fair-serving-objective}.
Effective electricity use concerns the service obtained from the budget, balanced throughput concerns sharing among concurrent requests, and demand-aware allocation concerns when electricity is spent.
Together, these properties connect the hourly fairness objective to the allocation decisions made during service.\looseness=-1

\noindent\textbf{Property 1: Effective electricity use.}
\phantomsection\label{prop:effective}
Effective electricity use means improving weak requests' service through better scheduling and spending useful remaining budget on further service improvements.
The commitment limits total consumption, so electricity saved by efficient execution creates opportunities to improve service for requests with low normalized throughput elsewhere in the hour.
We assess this property using the observed frontier between the tenth percentile (P10) of $q_i^H$ and absolute commitment deviation $d_h$, and report $\frac{E_h}{C_h}$ to distinguish underuse from overuse.
A higher P10 at no greater deviation, or a smaller deviation at no lower P10, represents more effective electricity use \mbox{among the compared operating points.}

\noindent\textbf{Property 2: Balanced throughput among concurrent requests.}
\phantomsection\label{prop:balanced}
Balanced throughput means that competing requests receive comparable service after accounting for standalone throughput and entitlement.
Concurrent requests share serving resources and electricity, so the throughput loss from contention should be shared according to entitlement.
We identify competition by positive overlap of arrival-to-completion or arrival-to-cutoff intervals, including queueing, and measure $r_{ij}=\frac{\min(q_i^H,q_j^H)}{\max(q_i^H,q_j^H)}$.
Larger $r_{ij}$ means closer normalized throughputs: the ratio equals one for equal positive values and zero when only one request produces output; pairs with two zero values are counted separately.
These ratios quantify normalized-throughput \mbox{balance among concurrent requests.}

\Needspace{225pt}
\begin{wrapfigure}{r}{0.47\textwidth}
\vspace{-4pt}
\centering
\captionsetup{justification=raggedright,singlelinecheck=false}
\includegraphics[width=\linewidth,trim=0 0 0 1pt,clip]{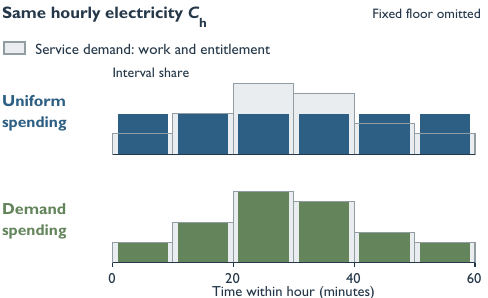}
\caption{Equal hourly electricity with uniform or demand-aware spending.
Demand-aware expenditure assigns larger shares to intervals with more work or higher entitlements.
The fixed operating floor is excluded from both profiles.}
\label{fig:temporal-illustration}
\vspace{-3pt}
\end{wrapfigure}
\noindent\textbf{Property 3: Demand-aware electricity allocation.}
\phantomsection\label{prop:demand-aware}
Demand-aware allocation assigns more adjustable electricity to periods with greater service needs, including more work or higher entitlements.
Uniform allocation can leave busy periods with insufficient electricity while reserving more electricity than quiet periods need; entitlements also affect the electricity required to reach $z$.
For evaluation, all methods use the same reference demand profile, determined by request arrivals, workloads, and entitlements at the common target $z$.
Let $D_k$ denote this reference service-electricity demand in interval $k$, and $A_k$ the actual electricity consumption in that interval after subtracting fixed operating costs.
We compare the hourly shares of demand and \mbox{expenditure using the total variation metric:}
\begin{equation}
\mathrm{TV}=\frac{1}{2}\sum_k
\left|\frac{D_k}{\sum_j D_j}-\frac{A_k}{\sum_j A_j}\right|.
\label{eq:tv}
\end{equation}

Lower TV indicates closer alignment: a TV of $0.02$ means that 2\% of adjustable electricity would need to be reallocated across intervals to match the demand shares.
Figure~\ref{fig:temporal-illustration} illustrates this timing choice; the appendix specifies the common reference-demand calculation.
Together with backlog and throughput by arrival time, TV shows whether expenditure follows service needs across the hour. \S~\ref{sec:design} develops the online scheduler for this objective and these properties.

\section{Scheduling Framework}
\label{sec:design}
\label{sec:design-overview}

\begin{figure*}[!htbp]
\centering
\includegraphics[width=\textwidth,trim=0 6pt 0 1.5pt,clip]{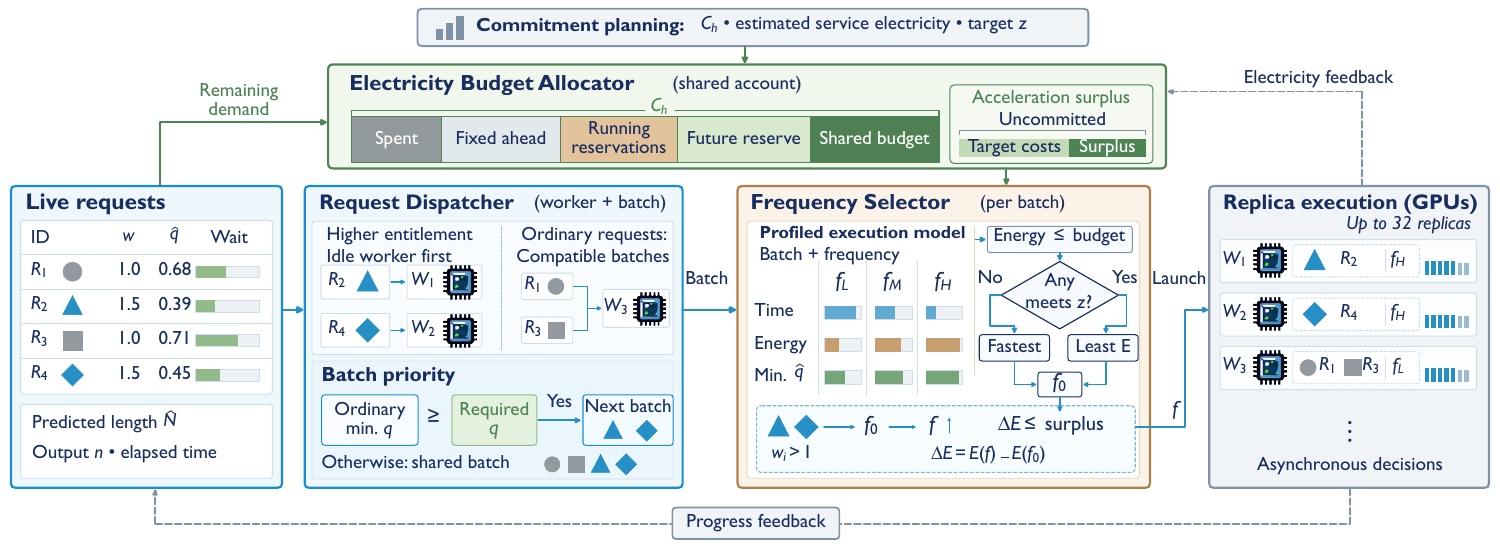}
\caption{\sys dispatches requests, allocates a shared electricity budget and surplus, and selects batch frequencies.
Required $q$ denotes the priority threshold (\S~\ref{sec:service-dispatch}); bars compare iteration time, electricity, and minimum predicted normalized throughput.
The surplus covers acceleration after target demand is reserved.
Launches reserve electricity; execution feedback updates request progress and consumption.}
\label{fig:design-overview}

\end{figure*}

\sys implements an online scheduling heuristic for the constrained fairness problem in Equation~\ref{eq:fair-serving-objective}.
On each arrival or iteration completion, \proc{Request Dispatcher} forms worker--batch assignments (\S~\ref{sec:service-dispatch}); \proc{Electricity Budget Allocator} determines the electricity available to those batches (\S~\ref{sec:electricity-spending}); and \proc{Frequency Selector} chooses each batch's GPU frequency (\S~\ref{sec:frequency-choice}).
\S~\ref{sec:execution-feedback} closes this decision loop with asynchronous execution and feedback.
Figure~\ref{fig:design-overview} and Algorithm~\ref{alg:online-scheduling} summarize how the three components turn the hourly budget into successive service opportunities.\looseness=-1

\begin{algorithm}[H]
\caption{Online scheduling with a shared budget}
\label{alg:online-scheduling}
\label{alg:service-dispatch}
\label{alg:frequency-choice}
\small
\renewcommand{\algorithmiccomment}[1]{\textcolor[gray]{0.40}{\footnotesize\itshape $\triangleright$~#1}}
\begin{algorithmic}[1]
\REQUIRE Hourly commitment $C_h$ and associated demand estimate; normalized-throughput target $z$; frequencies $\mathcal F$.
\STATE \textbf{On arrival or iteration completion:}
\STATE Update request progress and electricity account
\item[] \COMMENT{Assign workers and form batches (\S~\ref{sec:service-dispatch}).}
\item[] \COMMENT{$\mathcal P$: ready (worker $j$, batch $\mathcal B$) pairs.}
\STATE $\mathcal P\gets\proc{Request Dispatcher}()$
\FOR{each $(j,\mathcal B)\in\mathcal P$}
  \item[] \COMMENT{Budget current work and reserve for later (\S~\ref{sec:electricity-spending}).}
  \item[] \COMMENT{budget: shared electricity available for new iterations;}
  \item[] \COMMENT{surplus: shared electricity beyond target demand.}
  \STATE $(\mathit{budget},\mathit{surplus})\gets$
  \item[] \hspace{\algorithmicindent}\proc{Electricity Budget Allocator}()
  \item[] \COMMENT{Choose batch frequency $f$; use surplus to accelerate higher entitlement work (\S~\ref{sec:frequency-choice}).}
  \STATE $f\gets\proc{Frequency Selector}($
  \item[] \hspace{\algorithmicindent}$j,\mathcal B,\mathit{budget},\mathit{surplus})$
  \item[] \COMMENT{$\widehat E(\mathcal B,f)$: predicted incremental electricity for the next iteration.}
  \STATE Reserve $\widehat E(\mathcal B,f)$ in the shared account
  \STATE Launch batch $\mathcal B$ on worker $j$ at frequency $f$
\ENDFOR
\end{algorithmic}
\end{algorithm}

\subsection{\proc{Request Dispatcher}}
\label{sec:service-dispatch}

\proc{Request Dispatcher} controls worker assignment and batch participation to reduce waiting for higher entitlement requests while protecting ordinary service and preserving batching efficiency.
A batch contains the prefill and decoding work issued in one serving iteration; unfinished requests continue in later iterations as arrivals join available capacity.\looseness=-1

\noindent\textbf{Prediction and runtime state.}
Beyond the problem inputs in \S~\ref{sec:problem-overview}, \sys uses output-length estimates and a profiled execution model to compare candidate actions.
The profiled execution model uses offline GPU measurements to predict iteration duration and power from batch composition, context lengths, and frequency.
For each request that has arrived, a predicted output length $\widehat N_i$ can be supplied by a prompt-based predictor~\cite{jin2023s3}.
The profiled execution model then estimates standalone throughput $\widehat B_i$ for that length; Appendix~C describes the profiles and calibration checks.
\sys maintains observed output $n_i(t)$, elapsed waiting time, queue assignments, worker availability, and electricity reservations.
These estimates support comparisons of normalized throughput and electricity cost.\looseness=-1

\noindent\textbf{Normalized throughput during serving.}
\proc{Request Dispatcher} tracks normalized throughput to compare service progress while requests remain active.
At $t>a_i$, the online estimate is $\widehat q_i(t)=\frac{n_i(t)}{w_i\widehat B_i(t-a_i)}$, whose denominator includes queueing, prefill, and decoding.
After output begins, waiting without output lowers normalized throughput, while execution advances the numerator.
Before output begins, observed throughput is zero, so \sys predicts completion throughput from elapsed waiting, predicted length, and estimated standalone throughput.
Both cases include time already spent waiting when determining request priority.\looseness=-1

\noindent\textbf{Worker assignment.}
\proc{Request Dispatcher} consolidates ordinary requests ($w_i=1$) and gives admission preference to higher entitlement requests ($w_i>1$).
For higher entitlement arrivals, \sys prefers idle workers, then favors workers whose residents have the same entitlement, considering capacity and congestion.
A worker serving only ordinary requests can accept the arriving request when its resident requests can yield the next batch under the priority rule below and the new request can be admitted no later than it would be on the preferred worker.
For ordinary arrivals, \sys favors batches of ordinary requests but uses an idle worker when either a resident request's normalized throughput or the arriving request's predicted normalized throughput after waiting for admission falls below the median for active requests with higher entitlements that have produced output.
These rules combine earlier admission for higher entitlements with efficient batching of ordinary work.\looseness=-1

\noindent\textbf{Batch priority.}
\proc{Request Dispatcher} gives higher entitlement requests priority when ordinary requests' normalized throughput is sufficiently high.
On a shared worker, every ordinary resident's normalized throughput must reach both $z$ and the median normalized throughput of active higher entitlement requests that have produced output.
If this median is unavailable, the threshold is $z$; ordinary requests without output use their predicted normalized throughput at completion.
When the condition holds, higher entitlement requests receive priority for the next batch; otherwise, both classes remain eligible.
Batches obey token, sequence, and KV-memory limits, yielding ready worker--batch pairs $\mathcal P$.
These rules give higher entitlement requests additional service opportunities while retaining ordinary access to shared capacity; the resulting ready batches receive electricity allocations in \S~\ref{sec:electricity-spending}.\looseness=-1

\subsection{\proc{Electricity Budget Allocator}}
\label{sec:electricity-spending}

\proc{Electricity Budget Allocator} translates the hourly commitment into a shared budget for current work by accounting for past consumption, running iterations' reserved electricity, and expected later demand.

\noindent\textbf{Electricity reservations.}
The account deducts past consumption, future fixed costs, and electricity reserved for running work before allocating new service.
At time $t$ in hour $h$, let $E_h(t)$ denote realized consumption, $P_{\mathrm{floor}}$ the declared fixed operating power, $H$ the hour endpoint, and $\widehat E_{\mathrm{run}}(t)$ the incremental electricity still owed to running work.
The electricity remaining \mbox{for uncommitted service is}
\begin{equation}
R_h(t)={}C_h-E_h(t)-P_{\mathrm{floor}}(H-t)-\widehat E_{\mathrm{run}}(t).
\label{eq:remaining-electricity}
\end{equation}

Completion feedback reconciles reservations with consumption, preventing duplicate allocations across workers.

\noindent\textbf{Current and future demand.}
\proc{Electricity Budget Allocator} divides the remaining electricity between arrived work and later arrivals.
For arrived work, \sys estimates remaining demand using $z$, entitlements, predicted lengths, and observed output.
For later arrivals, \sys uses the aggregate estimate associated with $C_h$, accounting for requests already observed.
Reserving a proportional share of $R_h(t)$ for later arrivals leaves the current shared budget $A(t)$.
All current batches draw from this budget, with reserved work excluded from uncommitted demand.
Each launch reserves electricity before the next \mbox{allocation, so batches share one account.}

\noindent\textbf{Surplus electricity.}
The shared surplus $S(t)$ is the nonnegative electricity left after covering current and future service demand at target $z$.
At a higher GPU frequency, the predicted incremental electricity consumption of the next iteration must not exceed the current budget $A(t)$, and the additional electricity required for acceleration must not exceed the shared surplus $S(t)$.
These two checks permit acceleration while preserving the shared allocation; the appendix gives the budget formulas.
\S~\ref{sec:frequency-choice} uses the shared budget and surplus to select GPU \mbox{frequencies for each ready worker--batch pair.}

\subsection{\proc{Frequency Selector}}
\label{sec:frequency-choice}

\begingroup\parfillskip=0.4\linewidth minus 0.4\linewidth\relax For each batch, \proc{Frequency Selector} uses projected service and electricity cost to select a frequency for the common target, then considers additional acceleration for work with higher entitlements.\par\endgroup

\noindent\textbf{Throughput and electricity prediction.}
The profiled execution model introduced in \S~\ref{sec:service-dispatch} estimates execution costs at each candidate frequency.
For batch $\mathcal B$ and $f\in\mathcal F$, the model estimates iteration duration $\widehat\tau(\mathcal B,f)$ and incremental electricity $\widehat E(\mathcal B,f)$.
The output-length estimate $\widehat N_i$ and observed progress turn these execution estimates into average throughput $\widehat T_i(f)$ through completion or the hour endpoint, including elapsed waiting and output already produced.
Predicted normalized throughput is $\widehat q_i(f)=\frac{\widehat T_i(f)}{w_i\widehat B_i}$.
Decode-only predictions use batch speed and context costs; supported prefill predictions include remaining prefill and decoding, linking each candidate frequency to the request's projected service.\looseness=-1

\noindent\textbf{Meeting the service target.}
\proc{Frequency Selector} chooses the least costly affordable frequency predicted to meet the target for all batch members.
Affordability requires $\widehat E(\mathcal B,f)\le A(t)$; target feasibility requires a valid completion prediction with $\min_{i\in\mathcal B}\widehat q_i(f)\ge z$.
\sys compares next-iteration electricity for decode-only batches and predicted remaining service electricity for prefill batches of ordinary requests when that estimate is available.
The chosen frequency minimizes the applicable cost estimate; each launch reserves only the \mbox{electricity needed for the next iteration.}

\noindent\textbf{Fallback frequency selection.}
\proc{Frequency Selector} favors fast execution when no frequency can meet the target or completion predictions are unavailable.
\sys selects the frequency with the shortest predicted iteration time among those whose electricity fits $A(t)$.
If none fits, \sys minimizes predicted iteration electricity and records the expenditure.
One-iteration decisions let subsequent choices use updated progress and consumption.\looseness=-1

\noindent\textbf{Acceleration with surplus electricity.}
After reserving electricity for current and future service demand at target $z$ (\S~\ref{sec:electricity-spending}), \proc{Frequency Selector} uses the surplus $S(t)$ to accelerate higher entitlement batches.
Relative to the selected frequency $f_0$, a faster frequency must keep predicted iteration electricity within $A(t)$ and spend at most $S(t)$ more.
If $f_0$ meets the target, acceleration must preserve the target for every batch member.
If no faster frequency qualifies, \sys retains $f_0$.
Savings from batching and frequency selection thus support faster service for higher entitlement requests.
Execution feedback then updates the estimates and reservations used by the next decision (\S~\ref{sec:execution-feedback}).\looseness=-1

\subsection{Asynchronous Execution}
\label{sec:execution-feedback}

Execution feedback keeps service predictions and electricity reservations aligned with current progress.
The scheduler asynchronously prepares subsequent batches while already launched GPU work continues.
Arrivals and iteration completions update request progress, worker availability, and electricity reservations before triggering the next decision.
When an unfinished request reaches its predicted output length, its length estimate is extended; when it completes, its electricity-demand estimate is reconciled using the final output-token count.
\S~\ref{sec:length-robustness} evaluates sensitivity to length errors.
At iteration boundaries, workers admit arrivals, continue unfinished requests, and remove completed requests within capacity limits.
Each launch reserves electricity before another allocation, and scheduler response time affects subsequent launches.
These updates let subsequent decisions incorporate observed progress and consumption while preserving asynchronous execution.\looseness=-1

\FloatBarrier
\section{Evaluation}
\label{sec:evaluation}

We evaluate whether \sys improves fair LLM serving under hourly electricity commitments through three questions.
After describing the experimental setup (\S~\ref{sec:experimental-setup}), we evaluate the constrained fairness objective (\S~\ref{sec:constrained-fairness}), assess the three properties of fair serving (\S~\ref{sec:property-evaluation}), and examine scheduling efficiency and robustness (\S~\ref{sec:scheduler-evaluation}).
\begin{itemize}[leftmargin=*,itemsep=2pt,topsep=3pt,parsep=0pt,partopsep=0pt,before={\interlinepenalty=10000}]
\item \textbf{RQ1:} How does \sys improve the fairness objective while satisfying the hourly electricity commitment?
\item \textbf{RQ2:} How well does \sys achieve the following properties: effective electricity use, balanced throughput among concurrent requests, and demand-aware electricity \mbox{allocation? (Properties~1--3, \S~\ref{sec:properties}.)}
\item \textbf{RQ3:} How quickly does \sys make online scheduling decisions, and how robust is the resulting service \mbox{to output-length prediction errors?}
\end{itemize}

\subsection{Experimental Setup}
\label{sec:experimental-setup}

Methods share the same requests and hourly commitments.
We measure service using normalized throughput $q_i^H$ (\S~\ref{sec:fair-serving}): output produced by the hour's end divided by elapsed request time, standalone throughput (the output rate when served alone), and entitlement.
Elapsed time runs from arrival to completion, or to the hour boundary for unfinished requests; larger $q_i^H$ indicates better normalized service.\looseness=-1

\noindent\textbf{Runtime configuration.} We evaluate Qwen3.8-27B-FP8 with tensor parallelism of one (TP=1) in H20 and H200 configurations, each with 32 model replicas.\looseness=-1

\noindent\textbf{Workload traces.} The H20 configuration replays the Azure inference trace from May 12, 2024, 14:00 to 15:00 UTC (35,597 requests); H200 uses a heavier trace from May 11, 2024, 00:00 to 01:00 UTC (60,686 requests)~\cite{azure2024trace}.
Both use entitlements $w_i\in\{1,1.5\}$.
The main comparison supplies true recorded output lengths ($\widehat N_i=N_i$) and an accurate estimate of the total service-electricity demand used to set $C_h$.
\S~\ref{sec:length-robustness} evaluates errors in the per-request output-length estimates.\looseness=-1

\noindent\textbf{Hourly commitments.} All methods use the same electricity budget for each workload.
We set the normalized-throughput target $z=0.6$ before evaluating the policies, estimate the electricity needed for the request mix and entitlements, and add the pool's fixed operating cost.
The commitments are 17.68~MJ for H20 and 19.01~MJ for H200.
A configuration satisfies the commitment when its realized consumption obeys $E_h\le C_h$.\looseness=-1

\noindent\textbf{Baselines.} We compare methods that prioritize aggregate throughput, client fairness, or electricity efficiency.
\begin{itemize}[leftmargin=*,itemsep=2pt,topsep=3pt,parsep=0pt,partopsep=0pt,before={\interlinepenalty=10000}]
\item \textbf{vLLM} uses continuous batching and KV-cache management for efficient serving~\cite{kwon2023pagedattention}.
\item \textbf{Weighted VTC} prioritizes clients according to accumulated service relative to their entitlements~\cite{sheng2024vtc}.
We use costs derived from profiling to account for each client's accumulated service.
\item \textbf{EcoInfer} selects iteration frequencies to reduce electricity use while meeting its latency objective~\cite{hu2026ecoinfer}.
We retain that objective in the comparison.
\end{itemize}
vLLM and Weighted VTC do not account for hourly electricity commitments and exceed the budget at their default frequencies.
For RQ1, we retain their scheduling policies and select frequencies whose consumption is closest to the commitment without exceeding the commitment: 1365~MHz on H20 and 1170~MHz on H200.
The effective-electricity comparison in RQ2 also includes their default frequencies; all other baseline comparisons use configurations \mbox{that satisfy the commitment.}

\noindent\textbf{Evaluation metrics.} RQ1 compares the leximin objective and the distribution of normalized throughput.
Leximin sorts $q_i^H$ from lowest to highest and compares methods at the first rank where the values differ (Equation~\ref{eq:fair-serving-objective}).
We report the minimum $Q_{\min}$, zero-output counts, P10, and P1, where P10 and P1 are the tenth and first percentiles of $q_i^H$.
Each request has equal weight; a higher value at a given percentile indicates better normalized service.
RQ2 assesses P10 against absolute commitment deviation $d_h$ (\S~\ref{sec:commitment}), concurrent-request throughput ratios $r_{ij}$ (Property~2 in \S~\ref{sec:properties}), and temporal demand--electricity mismatch measured by total variation (TV; Equation~\ref{eq:tv}).
RQ3 measures scheduling response time and changes in request-level P10 under output-length errors.
Each property result explains the corresponding metric.\looseness=-1

\subsection{Fairness within the Hourly Commitment}
\label{sec:constrained-fairness}

\noindent\textbf{Leximin fairness.}
\sys improves the objective in Equation~\ref{eq:fair-serving-objective} over all baseline configurations satisfying the hourly commitment, answering RQ1 (Table~\ref{tab:main-results}).
All methods have the same minimum normalized throughput of zero because some requests arriving within the final 1.24~s on H20 or 0.96~s on H200 produce no output before the hour ends.
\sys leaves six and four such requests, respectively, versus at least seven and five for the baselines.
At the first differing rank, 7 on H20 and 5 on H200, \sys achieves normalized throughput of 0.1544 and 0.0158, while every baseline still has zero.
The sorted \sys normalized-throughput vector therefore ranks above each baseline under the leximin objective.\looseness=-1

\begin{table}[!htbp]
\centering
\captionsetup{width=.84\linewidth,justification=raggedright,singlelinecheck=false}
\caption{Normalized throughput under $E_h\le C_h$.
$Q_{\min}$ is the minimum, and Zero counts requests \mbox{without output at hour end.}}
\label{tab:main-results}
\small\renewcommand{\arraystretch}{1.09}\setlength{\tabcolsep}{5pt}
\begin{tabular*}{.80\linewidth}{@{\extracolsep{\fill}}llrrrr@{}}
\toprule
GPU & Method & $Q_{\min}$ & Zero & P10 $\uparrow$ & $\frac{E_h}{C_h}$ \\
\midrule
\textbf{H20} & \sys & 0 & \textbf{6} & \textbf{0.6299} & 0.9974 \\
& vLLM & 0 & 7 & 0.4783 & 0.9978 \\
& Weighted VTC & 0 & 7 & 0.4807 & 0.9971 \\
& EcoInfer & 0 & 13 & 0.3149 & 0.9376 \\
\midrule
\textbf{H200} & \sys & 0 & \textbf{4} & \textbf{0.6000} & 0.9998 \\
& vLLM & 0 & 5 & 0.4640 & 0.9700 \\
& Weighted VTC & 0 & 5 & 0.4686 & 0.9684 \\
& EcoInfer & 0 & 10 & 0.2021 & 0.8613 \\
\bottomrule
\end{tabular*}
\end{table}

\Needspace{235pt}
\begin{wrapfigure}{r}{0.49\textwidth}
\vspace{-5pt}
\centering
\captionsetup{justification=raggedright,singlelinecheck=false}
\includegraphics[width=\linewidth,trim=0 3pt 0 1.5pt,clip]{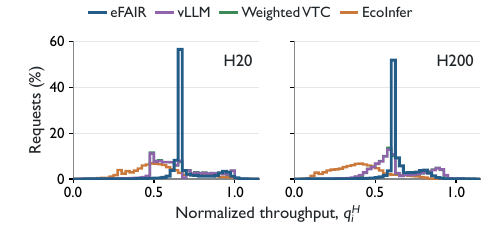}
\caption{Distribution of hour-end normalized throughput $q_i^H$ for H20 (left) and H200 (right).
Each curve shows the percentage of requests in 0.025-wide bins over $[0,1.15]$, with equal weight per request.
The distributions show higher lower-tail \mbox{normalized throughput under \sys.}}
\label{fig:request-distribution}
\end{wrapfigure}
\noindent\textbf{Service improvements for weak requests.}
\sys's leximin advantage is accompanied by higher lower-tail throughput and improvements for most of each baseline's weakest requests.
P10 improves by up to 197.0\% over the baselines that satisfy the commitment.
On H20, P10 increases by 31.7\% over vLLM and 31.0\% over Weighted VTC, with electricity differing by less than 0.04\%.
On H200, the gains are 29.3\% and 28.0\%, using 3.1\% to 3.2\% more electricity within the commitment.
The P10 gains over EcoInfer are 100.1\% and 197.0\%, respectively.
P1 also increases to 0.4649 on H20 and 0.4339 on H200, compared with baseline ranges of 0.1898--0.3229 and 0.1143--0.3263.
Figure~\ref{fig:request-distribution} further shows a concentration of requests at higher normalized throughput under \sys.
To determine whether requests poorly served by the baselines benefit, we select the 10\% of requests with the lowest $q_i^H$ under each baseline and compare those same requests under \sys.
Across these groups, \sys improves 3,522--3,549 of 3,560 requests on H20 and 5,989--6,043 of 6,069 on H200; detailed counts appear in the supplementary appendix.
Together, the percentile gains and comparisons of the same requests show that the improvement extends beyond the first differing rank and reaches requests with \mbox{the weakest baseline service.}\looseness=-1

\noindent\textbf{Explaining baseline performance.}
The results are consistent with differences in how the methods allocate service and control GPU frequency.
vLLM favors efficient batching without explicitly prioritizing requests with low normalized throughput.
Weighted VTC changes batch-admission order to balance accumulated client service relative to entitlement; when client service is already balanced, this ordering offers limited scope to improve individual request throughput.
The evaluated vLLM and Weighted VTC configurations also use the same fixed GPU frequency, consistent with their similar P10.
EcoInfer selects energy-saving frequencies to meet latency requirements, without explicitly using the remaining hourly budget to improve requests with low normalized throughput.
\sys coordinates batch participation and GPU frequency selection to direct the available electricity toward these requests.\looseness=-1

\subsection{Properties of Fair Serving}
\label{sec:property-evaluation}
\label{sec:effective-electricity}
\label{sec:concurrent-balance}
\label{sec:demand-allocation}

RQ2 evaluates the three properties in \S~\ref{sec:properties}: service obtained from the budget, throughput balance among concurrent requests, and electricity allocation across time.\looseness=-1

\Needspace{245pt}
\begin{wrapfigure}{l}{0.51\textwidth}
\vspace{-5pt}
\centering
\captionsetup{justification=raggedright,singlelinecheck=false}
\includegraphics[width=\linewidth,trim=0 3pt 0 0.5pt,clip]{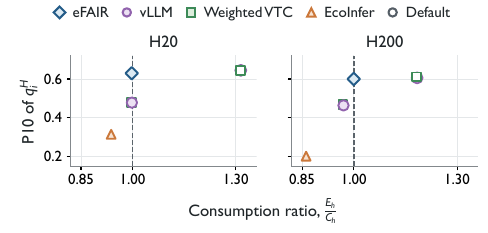}
\caption{P10 of normalized throughput versus hourly consumption relative to commitment for H20 (left) and H200 (right).
Higher P10 indicates better lower-tail service, and the vertical line at $\frac{E_h}{C_h}=1$ marks the electricity target.
Filled markers satisfy the commitment; hollow markers show default vLLM and Weighted VTC.
\sys approaches the defaults' P10 while \mbox{remaining within the commitment.}}
\label{fig:quality-electricity}
\end{wrapfigure}
\noindent\textbf{Effective electricity use.}
\sys achieves high lower-tail normalized throughput while keeping consumption close to the hourly commitment (\mbox{Figure~\ref{fig:quality-electricity}}).
For Property~1, higher P10 and lower absolute deviation $d_h$ are preferable; deviation decreases as $\frac{E_h}{C_h}$ approaches one.
\sys lies on the observed Pareto frontier: no evaluated configuration offers higher P10 without greater deviation, or lower deviation without lower P10.
The \sys deviations are 0.257\% on H20 and 0.017\% on H200.
Compared with EcoInfer, \sys increases P10 and reduces deviation.
On H20, vLLM has slightly lower deviation (0.221\%) but lower P10; on H200, \sys improves both metrics over all baselines \mbox{that satisfy the commitment.}
Default vLLM and Weighted VTC attain higher P10 but exceed the commitment by 31.4\% on H20 and 18.2\% to 18.4\% on H200.
Compared with these defaults, \sys retains 97.6\% to 97.7\% of P10 using 24.1\% less electricity on H20, and 98.2\% to 99.0\% using 15.4\% to 15.5\% less on H200.
These comparisons show that \sys approaches the defaults' lower-tail throughput while meeting the hourly commitment.\looseness=-1

\begin{figure*}[!htbp]
\centering
\includegraphics[width=\linewidth]{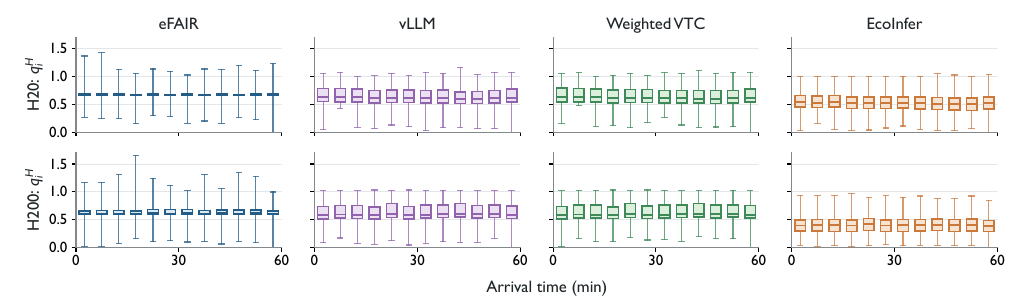}
\caption{Hour-end normalized throughput $q_i^H$ for requests grouped by five-minute arrival intervals.
Rows show H20 (top) and H200 (bottom); columns compare methods on the same arrival groups.
Boxes span P25--P75, center lines mark medians, and whiskers show the actual minimum and maximum.
\sys maintains similar group medians across the hour, indicating consistent \mbox{normalized service across arrival periods.}}
\label{fig:arrival-full-range}
\end{figure*}

\noindent\textbf{Balanced throughput among concurrent requests.}
\sys improves normalized-throughput balance among requests that compete for service.
For Property~2, $r_{ij}$ is the smaller of two concurrent requests' normalized throughputs divided by the larger; a ratio closer to one indicates more balanced service.
Table~\ref{tab:concurrent} reports the tenth percentile of these ratios, using the same request pairs with defined ratios under all four methods.
\sys improves P10 of $r_{ij}$ by up to 108.7\%, with the largest gain over EcoInfer on H200.
A separate comparison with an earlier \sys policy uses identical requests and commitments: request-level P10 increases by 8.86\% on H20 and 4.89\% on H200, while electricity increases by 0.77\% and 0.13\%.
The appendix details how these changes redistribute service from ordinary to higher entitlement requests.
These comparisons show improved concurrent balance and redistribution toward higher entitlements, alongside higher overall P10.\looseness=-1

\begin{table}[!htbp]
\centering
\captionsetup{width=.83\linewidth,justification=raggedright,singlelinecheck=false}
\caption{P10 of the normalized-throughput ratios $r_{ij}$ for request pairs concurrent under all four methods.
Higher values \mbox{indicate more balanced service.}}
\label{tab:concurrent}
\small\renewcommand{\arraystretch}{1.12}\setlength{\tabcolsep}{5pt}
\begin{tabular*}{.72\linewidth}{@{\extracolsep{\fill}}lrrrr@{}}
\toprule
GPU & \sys & vLLM & W. VTC & EcoInfer \\
\midrule
H20 & 0.7027 & 0.5636 & 0.5731 & 0.4490 \\
H200 & 0.7432 & 0.5825 & 0.5886 & 0.3561 \\
\bottomrule
\end{tabular*}
\end{table}

\noindent\textbf{Demand-aware electricity allocation.}
\sys follows service demand more closely and maintains consistent normalized throughput across arrival periods.
For Property~3, service-electricity demand is the reference electricity needed to meet the common service target, and adjustable electricity is actual consumption after fixed operating costs are subtracted.
TV compares their shares across five-minute intervals.
Lower TV indicates closer alignment, and zero means identical shares.
Table~\ref{tab:demand-alignment} reports TV of 0.001943 on H20 and 0.000472 on H200 for \sys.
The largest mismatch reduction is 95.9\%, attained against Weighted VTC on H200; the supplementary appendix plots the corresponding demand and electricity profiles.
The \proc{Electricity Budget Allocator} updates spending using remaining demand and commitment, whereas the baselines do not coordinate expenditure with both quantities.
Figure~\ref{fig:arrival-full-range} complements this temporal comparison by grouping hour-end normalized throughput according to request arrival time.
The \sys group medians span 0.6643--0.6652 on H20 and 0.6033--0.6178 on H200; boxes and whiskers show within-group variation.
The appendix also reports backlog and interval electricity measurements.
Lower TV and similar group medians together show that expenditure follows demand while service remains consistent across arrival periods.\looseness=-1

\begin{table}[!htbp]
\centering
\captionsetup{width=.75\linewidth,justification=raggedright,singlelinecheck=false}
\caption{Demand--electricity mismatch (TV; lower is better).}
\label{tab:demand-alignment}
\small\renewcommand{\arraystretch}{1.12}\setlength{\tabcolsep}{4pt}
\begin{tabular*}{.75\linewidth}{@{\extracolsep{\fill}}lrrrr@{}}
\toprule
GPU & \sys & vLLM & W. VTC & EcoInfer \\
\midrule
H20 & 0.001943 & 0.019165 & 0.019344 & 0.010561 \\
H200 & 0.000472 & 0.010192 & 0.011506 & 0.006716 \\
\bottomrule
\end{tabular*}
\end{table}

\subsection{Scheduling Efficiency and Robustness}
\label{sec:scheduler-evaluation}
\label{sec:control-overhead}
\label{sec:length-robustness}

RQ3 examines whether \sys responds promptly during serving and maintains normalized throughput when output-length predictions are inaccurate.\looseness=-1

\noindent\textbf{Scheduling response time.}
\sys produces decisions with millisecond-scale average response while GPU execution continues.
We measure response from the time a scheduling event enters the CPU queue until the resulting decision is available, including CPU processing and queueing during workload replay.
Table~\ref{tab:control-overhead} reports mean responses of 1.823~ms over 803,000 invocations on H20 and 1.609~ms over 1,264,780 invocations on H200.
The scheduler supplies each subsequent batch's configuration and electricity reservation before launch.
Scheduling updates thus overlap with already launched GPU work.\looseness=-1

\begin{table}[!htbp]
\centering
\captionsetup{width=.74\linewidth,justification=raggedright,singlelinecheck=false}
\caption{Scheduler invocations and enqueue-to-result response.}
\label{tab:control-overhead}
\small\renewcommand{\arraystretch}{1.10}\setlength{\tabcolsep}{6pt}
\begin{tabular*}{.69\linewidth}{@{\extracolsep{\fill}}lrr@{}}
\toprule
Metric & H20 & H200 \\
\midrule
Scheduler invocations & 803,000 & 1,264,780 \\
Mean response time (ms) & 1.823 & 1.609 \\
\bottomrule
\end{tabular*}
\end{table}

\noindent\textbf{Robustness to output-length prediction errors.}
\sys maintains higher P10 than all three baselines within the hourly commitment under the tested output-length errors.
Following the prediction-based VTC experiment, we compare true-length inputs ($\widehat N_i=N_i$) with four controlled perturbations: multiplying each length by a uniform draw from $[0.8,1.2]$ or $[0.5,1.5]$, or by a fixed factor of 0.8 or 1.2, then rounding to tokens.
These errors affect predicted completion times, per-request service-electricity demand, and online standalone-throughput estimates.
Actual requests, the fixed standalone throughputs $B_i$ used to calculate observed $q_i^H$, the commitment, and its associated aggregate demand estimate remain unchanged.
Relative to true-length inputs, P10 decreases by at most 0.81\% on H20 and 0.74\% on H200; absolute consumption deviations remain below 0.261\% and 0.018\%, respectively (Table~\ref{tab:length-robustness}).
The resulting service gains therefore persist across both random length errors and systematic overestimation or underestimation.\looseness=-1

\begin{table}[!htbp]
\centering
\captionsetup{width=.80\linewidth,justification=raggedright,singlelinecheck=false}
\caption{Length-error sensitivity: P10 and deviation $d_h$.}
\label{tab:length-robustness}
\small\renewcommand{\arraystretch}{1.1}\setlength{\tabcolsep}{5pt}
\begin{tabular*}{.79\linewidth}{@{\extracolsep{\fill}}lrrrr@{}}
\toprule
& \multicolumn{2}{c}{H20} & \multicolumn{2}{c}{H200} \\
\cmidrule(lr){2-3}\cmidrule(lr){4-5}
Prediction & P10 & $d_h$ (\%) & P10 & $d_h$ (\%) \\
\midrule
True lengths & 0.6302 & 0.246 & 0.6000 & 0.017 \\
Uniform $\pm20\%$ & 0.6308 & 0.260 & 0.5996 & 0.017 \\
Uniform $\pm50\%$ & 0.6298 & 0.253 & 0.5973 & 0.016 \\
$0.8\times$ length & 0.6404 & 0.182 & 0.6005 & 0.017 \\
$1.2\times$ length & 0.6251 & 0.215 & 0.5956 & 0.018 \\
\bottomrule
\end{tabular*}
\end{table}

\FloatBarrier
\section{Conclusion}
\label{sec:conclusion}

Fair LLM serving under hourly electricity commitments requires allocating both computing capacity among concurrent requests and electricity across arrival times.
We formulate this problem as leximin optimization of throughput normalized by standalone throughput and service entitlement, subject to the hourly electricity budget.
\sys uses a common service target to coordinate request dispatching, electricity reservations, and GPU frequency selection, directing available electricity toward requests with lower normalized throughput while accounting for future demand.
Across H20 and H200 configurations with $32$ replicas, \sys improves the tenth percentile of normalized throughput by $28.0\%$--$31.0\%$ over the best budget-feasible baseline, with absolute consumption deviations below $0.26\%$.
Additional results show more balanced normalized throughput among concurrent requests, closer alignment between electricity allocation and service demand, and \mbox{robustness to the tested output-length errors.}

\label{main-text-end}


\bibliographystyle{plainnat}
\bibliography{references}

\ifappendixdetails
\clearpage
\appendix
\numberwithin{equation}{section}
\numberwithin{table}{section}
\numberwithin{figure}{section}
\section{Commitment Planning and Electricity Demand}
\label{app:demand}

The common target $z$ connects the request throughput definition to both advance commitment planning and retrospective demand evaluation.
We first estimate the electricity required per request, then use these estimates to construct the hourly commitment and the common demand profile. The final accounting rules define the \mbox{electricity available for online decisions.}

\noindent\textbf{Per-request electricity demand.} We estimate the electricity needed to provide each request with normalized throughput of at least $z$.
For request $i$ and a fixed candidate configuration $c$, let $\widehat T_i(c)$ be the predicted average output throughput and $\widehat E_i(c)$ the incremental service electricity above the fixed operating floor.
The estimate is the minimum service electricity among candidate configurations that meet this target:
\begin{equation}
e_i(z)=\min_{c:\,\frac{\widehat T_i(c)}{w_iB_i}\ge z}\widehat E_i(c).
\label{eq:request-demand}
\end{equation}
Each estimate assumes that the request is served alone without queueing and includes prefill, decoding, and the execution gaps represented in the profile.
The target $z$, standalone throughput $B_i$, candidate configurations, and profiled execution model are fixed before method comparison.
If no configuration meets the target, the target is recorded as infeasible for that request. Every request in the two evaluated workloads has a feasible configuration.
Applying the same rule to every method gives a common reference \mbox{demand that accounts for entitlement.}

\noindent\textbf{Hourly commitment planning.} The commitment combines total expected service electricity with fixed operating electricity.
Let $\widehat D_h$ denote the estimated sum of $e_i(z)$ used in commitment planning for the expected hourly request and entitlement mix, and let $\Delta_h$ be \mbox{the hour duration. The commitment is}
\begin{equation}
C_h=P_{\mathrm{floor}}\Delta_h+\widehat D_h.
\label{eq:planned-commitment}
\end{equation}
The evaluated conditions use $z=0.6$ and an accurate aggregate service-demand estimate constructed from the recorded request population.
The \sys scheduler receives the aggregate estimate; individual requests become visible upon arrival.
H20 uses 7.468284~MJ of expected service electricity and 10.207262~MJ of fixed costs; H200 uses 8.463753~MJ and 10.544268~MJ, respectively.
The resulting commitment is fixed before evaluating any serving policy.

\noindent\textbf{Demand over time.} The common demand profile distributes each request's electricity estimate across the time required to meet the target.
For arrival time $a_i$ and total output count $N_i$, the target throughput $zw_iB_i$ gives the reference interval
\begin{equation}
J_i=\left[a_i,\ a_i+\frac{N_i}{w_iB_i z}\right].
\label{eq:target-interval}
\end{equation}
For five-minute interval $I_k$, the demand is
\begin{equation}
D_k=\sum_i e_i(z)\frac{|I_k\cap J_i|}{|J_i|}.
\label{eq:window-demand}
\end{equation}
Here $|\cdot|$ denotes interval duration, so a request crossing a boundary contributes only its overlapping fraction.
Recorded arrivals and request attributes instantiate this fixed rule; the target interval $J_i$ is identical across methods for each request.
All methods use the same $D_k$; reference demand beyond the hour is recorded outside the displayed hourly shares.
The calculation uses actual arrivals only after execution, while the online scheduler uses the aggregate service-demand estimate associated with $C_h$.

\noindent\textbf{Adjustable electricity.} Subtracting fixed operating costs from actual consumption isolates electricity consumed by serving work above the fixed operating floor.
For actual interval electricity $E_k$ and interval duration $|I_k|$,
\begin{equation}
A_k=E_k-P_{\mathrm{floor}}|I_k|.
\label{eq:adjustable-electricity}
\end{equation}
The fixed operating power is the pool's minimum GPU idle power plus the assumed scheduler operating power.
The scheduler power component is 500~W for \sys and 400~W for the baselines, while all methods use the same absolute $C_h$.
We retain total electricity $E_k$, demand $D_k$, and request backlog alongside the normalized shares used by TV.
This preserves the magnitude, timing, \mbox{and shape of electricity expenditure.}

\noindent\textbf{Current budget and surplus.} The \proc{Electricity Budget Allocator} reserves electricity for expected future requests before allocating the current budget.
Let $D_{\mathrm{future}}$ be the estimated service demand of later arrivals and $D_{\mathrm{active}}$ the demand of unfinished work not already reserved for execution.
For the unreserved electricity $R_h(t)$ defined in \mbox{the main paper, the current budget is}
\begin{equation}
A(t)=R_h(t)-\max\{R_h(t),0\}
\frac{D_{\mathrm{future}}}{D_{\mathrm{future}}+D_{\mathrm{active}}}.
\label{eq:demand-reservation}
\end{equation}
The reserved fraction is zero when the denominator is zero.
The surplus $S(t)$ is the positive part of $R_h(t)-D_{\mathrm{future}}-D_{\mathrm{active}}$.
Running work has already been deducted from $R_h(t)$ and from active uncommitted demand, preventing a second reservation for the same running work.
The current budget and surplus are shared across all \mbox{workers and updated after each reservation.}

\section{Evaluation Metrics and Aggregation}
\label{app:measurement}

All metrics use the recorded request IDs, service entitlements, and standalone throughputs for the evaluated workload.
We define the observation window, identify concurrent requests, and specify how request \mbox{and pair measurements are aggregated.}

\noindent\textbf{Observation window.} Every request arriving within the evaluated hour is included at the common endpoint.
For endpoint $H=3600$ seconds and request arrival $a_i<H$, let $n_i(H)$ be the produced output and define
\begin{equation}
\tau_i^H=\begin{cases}
c_i-a_i,&\text{completed by }H,\\
H-a_i,&\text{otherwise}.
\end{cases}
\label{eq:observed-duration}
\end{equation}
The request's normalized throughput is $q_i^H=\frac{n_i(H)}{w_iB_i\tau_i^H}$, using the same full-request standalone throughput $B_i$ for all policies.
Requests with no output receive zero normalized throughput, and service and electricity accounting both end at $H$.
This definition includes all observed output and elapsed waiting and execution time.

\noindent\textbf{Concurrent request pairs.} The main comparison uses the same request pairs under all four methods.
Request $i$ is outstanding on $[a_i,a_i+\tau_i^H)$.
The requests in each pair must overlap for a positive duration under every method; their ratios use hour-end normalized throughputs: $r_{ij}=\frac{\min(q_i^H,q_j^H)}{\max(q_i^H,q_j^H)}$.
If only one request has zero normalized throughput, the ratio is zero; if both values are zero, the ratio is undefined.
Pairs undefined under any method are excluded from all columns, retaining 922,870 of 922,941 pairs on H20 and 1,794,482 of 1,794,525 on H200.
Supplementary comparisons use separate \sys--baseline \mbox{intersections with explicit undefined counts.}

\noindent\textbf{Pair-level and request-level statistics.} Pair statistics give equal weight to concurrent pairs, while request statistics give equal weight to requests.
The main concurrent table takes the tenth percentile over defined pair ratios.
For the supplemental request summary, $f_i$ is the minimum defined $r_{ij}$ among request $i$'s competitors, and percentiles are then taken over requests with at least one defined competitor ratio.
The global minimum and requests without competitors are retained with the coverage counts.
These aggregation rules distinguish balance across concurrent pairs from each request's least balanced comparison.\looseness=-1

\noindent\textbf{Percentiles and matched requests.} Throughput percentiles and request-level comparisons give equal weight to each request.
For an ordered population of size $n$, quantiles use linear interpolation at zero-based index $(n-1)p$ for percentile fraction $p$.
The baseline weak group contains exactly $\lceil0.1n\rceil$ requests, ranked by baseline normalized throughput and then input order to resolve ties.
The same IDs are compared under \sys, counting higher, lower, and unchanged normalized throughput.
Arrival plots use the same five-minute arrival groups across methods; boxes report the 25th percentile, median, and 75th percentile, and whiskers retain the actual minimum and maximum.
These rules make all three types of service comparison reproducible.\looseness=-1

\section{Implementation and Measurement Details}
\label{app:execution}

This appendix describes the scheduler inputs and prediction procedures, the response-time and model measurements, and the configurations used for baseline comparisons.

\noindent\textbf{Scheduler inputs.} The \sys scheduler receives the hourly commitment and the aggregate service-electricity estimate used to set that commitment.
Individual requests become visible upon arrival; future request attributes and arrival times are not exposed.
The H20 population contains 12,065 ordinary requests and 23,532 with entitlement 1.5; H200 contains 20,900 and 39,786, respectively.
All methods and runs share these fixed input conditions.\looseness=-1

\noindent\textbf{Completion-time prediction.} Completion-time predictions combine the profiled execution model's batch estimates with the remaining prefill and decoding work.
Predicted output length $\widehat N_i$ and observed output determine remaining decoding work.
The fixed single-request reference profile sums per-output time costs through $\widehat N_i$ outputs to estimate standalone throughput $\widehat B_i$.
Evaluation uses the fixed standalone throughput $B_i$ for the full request and the actual output length.
For decoding-only batches, the prediction applies the current batch's slowdown to the profiled context-dependent costs and includes recent CPU decision time.
For supported ordinary-request prefill batches with completion predictions, frequency selection compares the electricity required to complete service; for decoding-only batches, selection compares incremental iteration electricity among candidates predicted to meet the target.
For batches without a completion projection, \proc{Frequency Selector} selects the frequency with the shortest predicted iteration duration among those whose predicted electricity fits the current shared budget.
If none fits, \proc{Frequency Selector} selects the frequency with minimum predicted iteration electricity.
Each decision reserves electricity for the next iteration; subsequent decisions incorporate observed progress.\looseness=-1

\noindent\textbf{Output-length prediction errors.}
The study in Section~5.4 of the paper compares true-length inputs ($\widehat N_i=N_i$) with four controlled error conditions.
The random conditions use one fixed seed (20261028), with a separate multiplier for each request derived from the request ID.
Predictions are rounded to the nearest positive token count and bounded by the declared output and context limits; the public output cap is 8,192 for all requests and is independent of their realized lengths.
When an unfinished request reaches the predicted length, the estimate doubles, subject to these same limits.
Completion reveals the final output count and settles the request's reference work and arrived electricity demand.
The predicted length is used to estimate completion time, electricity demand, and standalone throughput.
Evaluation always uses the original full-request standalone throughput, commitment, arrival sequence, and actual output lengths.
The aggregate service-demand estimate used to set $C_h$ is held fixed to isolate per-request length uncertainty.
Ten full-hour runs cover five conditions on each hardware configuration, including matched true-length controls; each run uses a distinct physical CPU core.
Prediction errors, estimate extensions, and electricity deviations are archived.\looseness=-1

\noindent\textbf{Scheduling response-time measurement.}
The selected H20 run records 803,000 scheduler invocations and 790,499 serving iterations; H200 records 1,264,780 invocations and 1,242,954 iterations.
Invocation counts include calls that return without launching a batch.
The scheduling queue records enqueue, processing-start, and result-ready times.
CPU processing duration is timed around each scheduler callback; queueing and result-ready times follow the replay's event scheduler.
Mean response time is the sum of total queue wait and total scheduler processing time, divided by the invocation count; both totals use this same population.
On H20, mean queueing and processing times are 1.132487~ms and 0.690336~ms, giving a mean response time of 1.822823~ms.
On H200, they are 0.714393~ms and 0.895062~ms, giving a mean response time of 1.609455~ms.
The accumulated scheduler processing times, 554.34 and 1,132.06 seconds, exclude CPU queueing time.
Maximum recorded queue waits are 842.33~ms and 1,136.84~ms.
The replay overlaps serving iterations with scheduler processing; subsequent launches wait for the scheduling decision and the host/device launch constraints.
The reported implementation evaluates five frequency candidates sequentially.
The evaluations share dispatch, state-update, budget, and selection work.
Parallel candidate evaluation retains this shared work and adds task coordination.\looseness=-1

\noindent\textbf{Model calibration and validation.}
Hardware energy measurements use differences in NVML cumulative GPU energy counters over each profiling window.
The profiled execution model is checked against measurement records that were available during model development but excluded from fitting the correction parameters.
H20 duration correction uses single-request decode at 1980~MHz: 9,254 steps in the first 60 seconds are used to fit the correction, and 11,613 steps in the next 60 seconds are used to check it.
The reported timing error is $\frac{\sum_j|\widehat{t}_j-t_j|}{\sum_j t_j}$, where $t_j$ is measured step duration.
The timing error decreases from 2.6976\% to 2.1759\%; signed total-duration bias decreases from $-2.1549\%$ to $-0.0115\%$.
For effective decode power, batch-size-one windows are used to fit the correction and batch-size-two windows are used to check it at 345, 750, 1170, 1575, and 1980~MHz.
Each check contains 128 target calls with contexts from 3,080 to 3,209 tokens; corrected errors range from $-0.033\%$ to $+0.980\%$.
The duration check covers single-request decode at 1980~MHz, while the power checks cover the five listed frequencies.
The H200 power correction uses two disjoint intervals on eight GPUs from one 90-second capture.
Three separate saved measurement checks last 93.8, 101.0, and 305.7 seconds.
The first two use two-request, 2,048-token-prompt configurations at 1575 and 750~MHz; the third uses a Poisson arrival trace, mostly at 1170~MHz.
Their aggregate energy errors are $+2.22\%$, $-0.88\%$, and $-2.94\%$.
Counter sampling and window interpolation contribute uncertainty to these checks.\looseness=-1

\noindent\textbf{Worker assignment and batch selection.}
The \proc{Request Dispatcher} first uses an idle worker for a higher entitlement arrival; otherwise the dispatcher favors residents with the same entitlement, ready workers, and smaller groups, followed by earlier completion among busy workers.
Ordinary arrivals favor groups of ordinary requests below the configured batch-size limit, using ready groups before waiting for a busy one.
Higher entitlement requests receive batch priority when every ordinary resident reaches both $z$ and the median normalized throughput of active higher entitlement requests with output; without that median, the threshold is $z$.
Before output begins, the resident request's prediction uses elapsed time plus remaining service at the estimated standalone throughput.
The dispatcher uses an available idle worker when a destination resident or the incoming request falls below the higher entitlement median; without that median the dispatcher retains the preferred destination.
The incoming request estimate includes the selected worker's remaining execution before admission.
Token, sequence, and memory limits determine whether a batch is feasible.\looseness=-1

\noindent\textbf{Baseline configurations.}
The principal vLLM and Weighted VTC operating points are 1365~MHz for H20 and 1170~MHz for H200; their default points use 1980 and 1710~MHz, respectively.
The same profiled VTC service cost applies at both operating points.
The H20 1365~MHz electricity and duration estimates use interpolation between the profiled 1170 and 1575~MHz calibration points.
Baseline frequencies are chosen to make consumption as close as possible to the commitment without exceeding the budget or changing the original scheduling policy.\looseness=-1

\section{Supplementary Results}
\label{app:results}

The supplementary results examine lower throughput percentiles, concurrent-request comparisons, electricity allocation over time, and earlier scheduling policies.
Tables identify their comparison populations; earlier policies are compared at the current commitment \mbox{unless a different commitment is stated.}

\begin{table}[!htbp]
\centering
\caption{Lower normalized-throughput percentiles and hour-end request counts.
$Q_{\min}=0$ for all methods because some requests have no output.}
\label{tab:tails}
\vspace{3pt}
\small\renewcommand{\arraystretch}{1.09}\setlength{\tabcolsep}{5pt}
\begin{tabular*}{.84\linewidth}{@{\extracolsep{\fill}}llrrrr@{}}
\toprule
GPU & Method & P1 & P0.1 & Zero & Unfinished \\
\midrule
H20 & \sys & 0.4649 & 0.2775 & 6 & 30 \\
H20 & vLLM & 0.3050 & 0.1304 & 7 & 29 \\
H20 & Weighted VTC & 0.3229 & 0.1327 & 7 & 28 \\
H20 & EcoInfer & 0.1898 & 0.0732 & 13 & 36 \\
H200 & \sys & 0.4339 & 0.1448 & 4 & 25 \\
H200 & vLLM & 0.3190 & 0.1833 & 5 & 23 \\
H200 & Weighted VTC & 0.3263 & 0.1511 & 5 & 23 \\
H200 & EcoInfer & 0.1143 & 0.0587 & 10 & 33 \\
\bottomrule
\end{tabular*}
\end{table}

\begin{figure}[!htbp]
\centering\includegraphics[width=.69\linewidth]{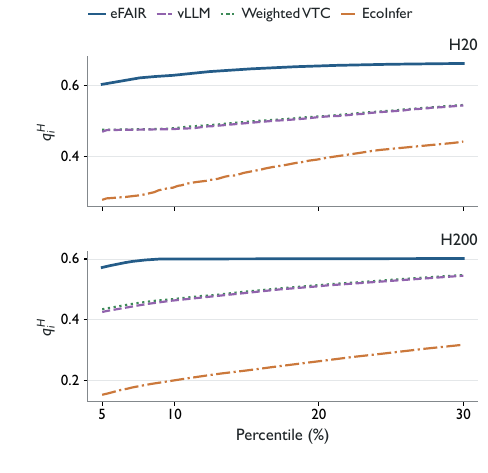}
\caption{Normalized throughput from P5 to P30 for H20 (top) and H200 (bottom). Higher values indicate better service at the same percentile. These curves show how the \mbox{lower-throughput requests compare across methods.}}
\label{fig:lower-quantiles}
\end{figure}

\noindent\textbf{Lower throughput percentiles.} Deeper percentiles show where the service gains are concentrated.
Table~\ref{tab:tails} reports P1, P0.1, zero output, and unfinished counts for all principal methods.
\sys has 6 requests with zero output and 30 unfinished requests on H20, and 4 and 25 on H200.
The P1 of \sys exceeds the baseline values in both configurations.
On H20, P0.1 also improves; on H200, the value of 0.1448 is below vLLM's 0.1833 and Weighted VTC's 0.1511.
The strict minimum remains zero for every method.
Figure~\ref{fig:lower-quantiles} shows the surrounding distribution so that improvements in the reported P10 summary are read alongside the tail.\looseness=-1

\begin{table}[!htbp]
\centering
\caption{Changes for each baseline's fixed set of weakest requests (10\%).
The counts compare normalized throughput for the same requests under each baseline and \sys. Ties include requests without output under both methods.}
\label{tab:weak}
\vspace{3pt}
\small\renewcommand{\arraystretch}{1.09}\setlength{\tabcolsep}{4pt}
\begin{tabular*}{.84\linewidth}{@{\extracolsep{\fill}}llrrrr@{}}
\toprule
GPU & Baseline & Count & Improved & Lower & Tied \\
\midrule
H20 & vLLM & 3,560 & 3,522 & 32 & 6 \\
H20 & Weighted VTC & 3,560 & 3,528 & 26 & 6 \\
H20 & EcoInfer & 3,560 & 3,549 & 5 & 6 \\
H200 & vLLM & 6,069 & 5,999 & 66 & 4 \\
H200 & Weighted VTC & 6,069 & 5,989 & 76 & 4 \\
H200 & EcoInfer & 6,069 & 6,043 & 22 & 4 \\
\bottomrule
\end{tabular*}
\end{table}

\noindent\textbf{Leximin comparison.} The empirical leximin comparison sorts normalized throughput for all 35,597 H20 requests or all 60,686 H200 requests, including zeros.
The first six H20 values and first four H200 values tie at zero across methods.
At the next rank, \sys is positive while every baseline remains at zero, establishing the ordering reported in the main paper.
The remaining sorted values describe the breadth of the gains: on H20, every sorted position through the lowest 10\% is at least as high under \sys; on H200, values at ranks 6--115 are lower than those under vLLM, and values at ranks 48--77 are lower than those under Weighted VTC.
The P1 and P10 values complement these local differences with stronger lower-tail service at larger ranks.\looseness=-1

\noindent\textbf{Matched-request comparison.} Most requests in each baseline's lowest-throughput decile receive higher normalized throughput under \sys.
Across the three baselines, \sys improves 3,522--3,549 of 3,560 H20 requests and 5,989--6,043 of 6,069 H200 requests.
Table~\ref{tab:weak} reports improvements, regressions, and ties for each baseline's fixed weakest 10\% request set.
The gains therefore reach the same weak requests across methods.\looseness=-1

\begin{table}[!htbp]
\centering
\captionsetup{width=.84\linewidth,justification=raggedright,singlelinecheck=false}
\caption{P10 of each request's worst competitor ratio $f_i$ for concurrent pairs shared by \sys and each baseline.
Higher is better.}
\label{tab:worst-competitor}
\vspace{3pt}
\small\renewcommand{\arraystretch}{1.09}\setlength{\tabcolsep}{5pt}
\begin{tabular*}{.70\linewidth}{@{\extracolsep{\fill}}llrr@{}}
\toprule
GPU & Baseline & \sys & Baseline \\
\midrule
H20 & vLLM & 0.5109 & 0.3142 \\
H20 & Weighted VTC & 0.5081 & 0.3361 \\
H20 & EcoInfer & 0.5021 & 0.2199 \\
H200 & vLLM & 0.5589 & 0.3587 \\
H200 & Weighted VTC & 0.5581 & 0.3611 \\
H200 & EcoInfer & 0.5446 & 0.1572 \\
\bottomrule
\end{tabular*}
\end{table}

\begin{table}[!htbp]
\centering
\caption{Concurrent-pair counts for supplementary \sys--baseline comparisons.
Zero/zero columns list undefined pairs for \sys and the baseline, respectively.
The last column counts shared pairs with a defined ratio under both methods.}
\label{tab:support}
\vspace{3pt}
\small\renewcommand{\arraystretch}{1.09}\setlength{\tabcolsep}{4pt}
\begin{tabular*}{.84\linewidth}{@{\extracolsep{\fill}}llrrrr@{}}
\toprule
GPU & Baseline & Shared & \multicolumn{2}{c}{Zero/zero} & Defined \\
\midrule
H20 & vLLM & 932,585 & 15 & 21 & 932,564 \\
H20 & Weighted VTC & 934,553 & 15 & 21 & 934,532 \\
H20 & EcoInfer & 949,985 & 15 & 71 & 949,914 \\
H200 & vLLM & 1,806,205 & 6 & 10 & 1,806,195 \\
H200 & Weighted VTC & 1,810,378 & 6 & 10 & 1,810,368 \\
H200 & EcoInfer & 1,928,404 & 6 & 43 & 1,928,361 \\
\bottomrule
\end{tabular*}
\end{table}

\noindent\textbf{Concurrent-request comparisons.} The supplementary comparison summarizes each request's least balanced concurrent pair.
Table~\ref{tab:worst-competitor} reports P10 of $f_i$, the minimum defined throughput ratio between request $i$ and any concurrent request (Appendix~\ref{app:measurement}), supplementing the pair-weighted statistic in the main paper.
Table~\ref{tab:support} lists the number of concurrent pairs shared by each compared method pair and the counts with undefined zero/zero ratios.
The H20 condition has one request without a competing request; all H200 requests have competitors.
The counts specify how many concurrent pairs contribute to each comparison.\looseness=-1

\begin{figure}[!htbp]
\centering
\includegraphics[width=.87\linewidth,trim=0 2pt 0 1.5pt,clip]{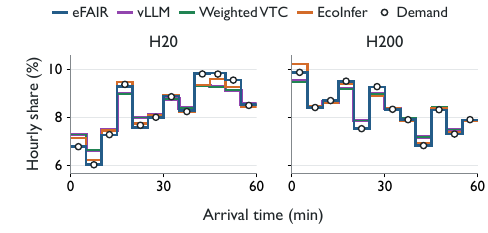}
\caption{Five-minute shares of reference electricity demand and adjustable electricity for H20 (left) and H200 (right).
Fixed operating costs are removed from electricity consumption, and each series is normalized by its own hourly total.
The \sys expenditure profile follows the common demand profile most closely, consistent with the TV values reported in the main paper.}
\label{fig:demand-electricity}
\end{figure}

\begin{figure}[!htbp]
\centering\includegraphics[width=.72\linewidth]{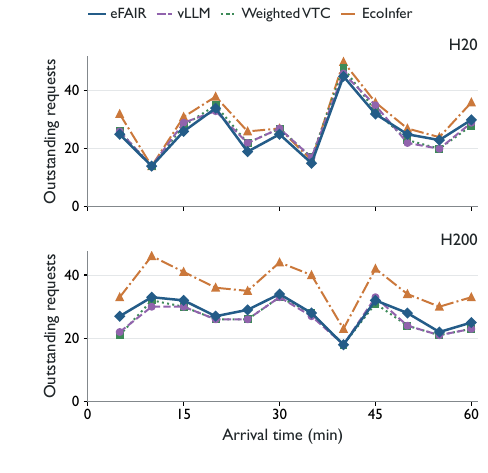}
\caption{Request backlog at five-minute boundaries, counting arrived requests that have not completed. The final point gives the unfinished population at the hour boundary. These curves place the electricity profiles in the context of outstanding service demand.}
\label{fig:backlog}
\end{figure}

\noindent\textbf{Request backlog.} Backlog places temporal electricity shares in the context of unfinished work.
Figure~\ref{fig:backlog} counts arrived requests that have not completed at each boundary.
Figure~\ref{fig:demand-electricity} uses the same time intervals for electricity accounting; the main paper's box plots group requests by their original arrival times.
The final backlog equals the unfinished counts in Table~\ref{tab:tails}.
Together, these views relate electricity expenditure to request arrivals, \mbox{normalized throughput, and unfinished work.}\looseness=-1

\begin{table}[!htbp]
\centering
\caption{The preceding policy and \sys at identical current inputs and commitments.
Electricity is total consumption.
The last two columns report P10 of normalized \mbox{throughput within each entitlement group.}}
\label{tab:controlled-policy}
\vspace{3pt}
\small\renewcommand{\arraystretch}{1.09}\setlength{\tabcolsep}{4pt}
\begin{tabular*}{.84\linewidth}{@{\extracolsep{\fill}}llrrrr@{}}
\toprule
GPU & Policy & $E_h$ (MJ) & All P10 & $w=1$ & $w=1.5$ \\
\midrule
H20 & Earlier & 17.496149 & 0.5786 & 0.6777 & 0.5522 \\
H20 & \sys & 17.630081 & 0.6299 & 0.6222 & 0.6416 \\
H200 & Earlier & 18.980518 & 0.5721 & 0.7558 & 0.5552 \\
H200 & \sys & 19.004839 & 0.6000 & 0.6102 & 0.5979 \\
\bottomrule
\end{tabular*}
\end{table}

\noindent\textbf{Comparison with an earlier policy.} At the same commitment, \sys redistributes service toward higher entitlement requests relative to the earlier policy.
Table~\ref{tab:controlled-policy} keeps the earlier policy at the current budget and request population.
\sys raises the aggregate P10 from 0.578642 to 0.629923 on H20 and from 0.572064 to 0.600043 on H200.
For higher entitlement requests ($w_i=1.5$), P10 rises from 0.5522 to 0.6416 on H20 and from 0.5552 to 0.5979 on H200; ordinary-request P10 falls by 8.18\% and 19.27\%, respectively.
Total electricity rises by 0.77\% and 0.13\%, respectively, alongside these service gains.\looseness=-1

\noindent\textbf{Earlier commitments.} Earlier experiments used different hourly commitments, so their P10 values correspond to different electricity constraints.
The previous H20 operating point used 18.583465~MJ and obtained P10 of 0.636222; the current point uses a smaller commitment and achieves P10 of 0.629923.
The previous H200 point used 18.001978~MJ with P10 of 0.523514; the current commitment is 5.59\% larger.
The original conditions scaled from default vLLM consumption and their raw outcomes remain in the supplied archive.
The current tables \mbox{use the new commitments consistently.}\looseness=-1

\noindent\textbf{Electricity demand and consumption by interval.} Absolute interval totals complement the normalized demand and electricity shares.
Tables~\ref{tab:h20-intervals} and~\ref{tab:h200-intervals} report the common demand $D_k$ and actual total electricity $E_k$ in each interval.
The method columns include the fixed cost; subtracting the fixed operating cost gives the adjustable electricity used by TV.
The demand totals cover only the reference service intervals that overlap the reported hour.
These values preserve the absolute expenditure alongside the normalized shares in Figure~\ref{fig:demand-electricity}.\looseness=-1

\begin{table}[!htbp]
\centering
\caption{H20 interval demand and total electricity, in kJ.
Intervals are in minutes.
Method columns report $E_k$, including \mbox{fixed costs; VTC denotes Weighted VTC.}}
\label{tab:h20-intervals}
\vspace{3pt}
\small\renewcommand{\arraystretch}{1.08}\setlength{\tabcolsep}{4pt}
\begin{tabular*}{.86\linewidth}{@{\extracolsep{\fill}}lrrrrr@{}}
\toprule
Interval & $D_k$ & \sys & vLLM & VTC & EcoInfer \\
\midrule
0 to 5 & 506.14 & 1354.84 & 1389.56 & 1389.07 & 1302.09 \\
5 to 10 & 450.36 & 1300.76 & 1337.99 & 1338.24 & 1241.32 \\
10 to 15 & 543.35 & 1391.25 & 1403.73 & 1402.73 & 1322.45 \\
15 to 20 & 698.40 & 1540.86 & 1521.56 & 1518.66 & 1457.07 \\
20 to 25 & 572.31 & 1414.70 & 1443.30 & 1442.92 & 1342.39 \\
25 to 30 & 596.90 & 1444.06 & 1452.98 & 1453.47 & 1366.24 \\
30 to 35 & 660.32 & 1510.36 & 1504.02 & 1502.48 & 1420.06 \\
35 to 40 & 614.53 & 1466.79 & 1474.64 & 1474.73 & 1374.32 \\
40 to 45 & 730.53 & 1580.14 & 1546.66 & 1545.28 & 1448.51 \\
45 to 50 & 730.32 & 1581.94 & 1543.18 & 1541.78 & 1465.80 \\
50 to 55 & 711.67 & 1560.09 & 1531.17 & 1531.81 & 1444.20 \\
55 to 60 & 633.48 & 1484.29 & 1487.65 & 1483.85 & 1387.32 \\
\midrule
Total & 7448.32 & 17630.08 & 17636.45 & 17625.04 & 16571.76 \\
\bottomrule
\end{tabular*}
\end{table}

\begin{table}[!htbp]
\centering
\caption{H200 interval demand and total electricity, in kJ.
Intervals are in minutes.
Method columns report $E_k$, including \mbox{fixed costs; VTC denotes Weighted VTC.}}
\label{tab:h200-intervals}
\vspace{3pt}
\small\renewcommand{\arraystretch}{1.08}\setlength{\tabcolsep}{4pt}
\begin{tabular*}{.86\linewidth}{@{\extracolsep{\fill}}lrrrrr@{}}
\toprule
Interval & $D_k$ & \sys & vLLM & VTC & EcoInfer \\
\midrule
0 to 5 & 835.45 & 1714.28 & 1635.78 & 1627.87 & 1480.52 \\
5 to 10 & 712.23 & 1591.91 & 1543.57 & 1541.69 & 1373.60 \\
10 to 15 & 736.63 & 1615.23 & 1562.49 & 1557.57 & 1382.20 \\
15 to 20 & 804.64 & 1683.09 & 1608.00 & 1604.43 & 1431.13 \\
20 to 25 & 638.03 & 1518.12 & 1499.14 & 1497.44 & 1315.31 \\
25 to 30 & 785.42 & 1662.96 & 1592.13 & 1586.84 & 1399.05 \\
30 to 35 & 706.01 & 1584.08 & 1540.84 & 1540.59 & 1367.24 \\
35 to 40 & 669.50 & 1548.12 & 1505.31 & 1504.50 & 1334.56 \\
40 to 45 & 578.69 & 1458.49 & 1442.96 & 1441.12 & 1277.40 \\
45 to 50 & 704.48 & 1582.62 & 1541.19 & 1542.01 & 1367.92 \\
50 to 55 & 619.32 & 1498.56 & 1466.98 & 1466.76 & 1306.71 \\
55 to 60 & 669.65 & 1547.37 & 1498.98 & 1496.62 & 1336.02 \\
\midrule
Total & 8460.03 & 19004.84 & 18437.38 & 18407.42 & 16371.65 \\
\bottomrule
\end{tabular*}
\end{table}

\begin{table}[!htbp]
\centering
\caption{Service and electricity at the common commitments.
vLLM and Weighted VTC use the selected frequencies.
\mbox{Higher P10 and lower $|\delta_h|$ and TV are better.}}
\label{tab:electricity-diagnostics}
\vspace{3pt}
\small\renewcommand{\arraystretch}{1.09}\setlength{\tabcolsep}{5pt}
\begin{tabular*}{.77\linewidth}{@{\extracolsep{\fill}}lrrrr@{}}
\toprule
Method & P10 $\uparrow$ & $\frac{E_h}{C_h}$ & $\delta_h$ (\%) & TV $\downarrow$ \\
\midrule
\multicolumn{5}{l}{\textbf{H20} ($C_h=17.675546$ MJ)} \\
\sys & 0.6299 & 0.9974 & -0.257 & 0.001943 \\
vLLM & 0.4783 & 0.9978 & -0.221 & 0.019165 \\
Weighted VTC & 0.4807 & 0.9971 & -0.286 & 0.019344 \\
EcoInfer & 0.3149 & 0.9376 & -6.245 & 0.010561 \\
\midrule
\multicolumn{5}{l}{\textbf{H200} ($C_h=19.008021$ MJ)} \\
\sys & 0.6000 & 0.9998 & -0.017 & 0.000472 \\
vLLM & 0.4640 & 0.9700 & -3.002 & 0.010192 \\
Weighted VTC & 0.4686 & 0.9684 & -3.160 & 0.011506 \\
EcoInfer & 0.2021 & 0.8613 & -13.870 & 0.006716 \\
\bottomrule
\end{tabular*}
\end{table}

\fi

\end{document}